\documentclass[sigconf]{acmart}

\usepackage{aecompl}

\AtBeginDocument{%
  \providecommand\BibTeX{{%
    \normalfont B\kern-0.5em{\scshape i\kern-0.25em b}\kern-0.8em\TeX}}}

\usepackage{amsmath,amsfonts,bm}

\def\eqref#1{equation~\ref{#1}}
\def\1{\bm{1}}

\def\ve{{\mathbf{e}}}

\def\vh{{\mathbf{h}}}

\def\vp{{\mathbf{p}}}
\def\vq{{\mathbf{q}}}

\def\vx{{\mathbf{x}}}

\def\mW{{\mathbf{W}}}

\DeclareMathAlphabet{\mathsfit}{\encodingdefault}{\sfdefault}{m}{sl}
\SetMathAlphabet{\mathsfit}{bold}{\encodingdefault}{\sfdefault}{bx}{n}
\newcommand{\tens}[1]{\mathbf{\mathsfit{#1}}}

\def\tO{{\tens{O}}}

\def\tR{{\tens{R}}}

\def\gA{{\mathcal{A}}}

\def\gL{{\mathcal{L}}}

\def\gN{{\mathcal{N}}}

\def\gP{{\mathcal{P}}}

\def\gR{{\mathcal{R}}}

\def\gT{{\mathcal{T}}}

\def\sR{{\mathbb{R}}}

\usepackage{balance}
\usepackage{hyperref}       % hyperlinks
\usepackage{url}            % simple URL typesetting
\usepackage{booktabs}       % professional-quality tables
\usepackage{amsfonts}       % blackboard math symbols
\usepackage{nicefrac}       % compact symbols for 1/2, etc.
\usepackage{microtype}      % microtypography
\usepackage{hyperref}
\usepackage{graphicx}
\usepackage{multirow}
\usepackage{algorithm}
\usepackage{algorithmic}
\usepackage{subfigure}

\copyrightyear{2021}
\acmYear{2021}
\setcopyright{acmcopyright}
\acmConference[CIKM '21]{Proceedings of the 30th ACM
International Conference on Information and Knowledge Management}{November
1--5, 2021}{Virtual Event, QLD, Australia}
\acmBooktitle{Proceedings of the 30th ACM International Conference on Information
and Knowledge Management (CIKM '21), November 1--5, 2021, Virtual Event, QLD,
Australia}
\acmPrice{15.00}
\acmDOI{10.1145/3459637.3482454}
\acmISBN{978-1-4503-8446-9/21/11}
\begin{document}
\fancyhead{}

\title{Neural PathSim for Inductive Similarity Search \\in Heterogeneous Information Networks
}

\author{Wenyi Xiao}
\affiliation{%
  \institution{Department of CSE, HKUST, Hong Kong, China}
}
\email{wxiaoae@cse.ust.hk}

\author{Huan Zhao}\authornote{Huan Zhao is the corresponding author.}
\affiliation{%
  \institution{4Paradigm Inc., Beijing, China}
}
\email{zhaohuan@4paradigm.com}

\author{Vincent W. Zheng}
\affiliation{%
  \institution{WeBank, Shenzhen, China}
}
\email{vincentz@webank.com}

\author{Yangqiu Song}
\affiliation{%
  \institution{Department of CSE, HKUST, Hong Kong, China}
}
\affiliation{%
  \institution{
  Peng Cheng Laboratory, Shenzhen, China}
}
\email{yqsong@cse.ust.hk}

\begin{abstract}
PathSim is a widely used meta-path-based similarity in heterogeneous information networks. 
Numerous applications rely on the computation of PathSim, including similarity search and clustering. Computing PathSim scores on large graphs is computationally challenging due to its high time and storage complexity. 
In this paper, we propose to transform the problem of approximating the ground truth PathSim scores into a learning problem.
We design an encoder-decoder based framework, NeuPath, where the algorithmic structure of PathSim is considered. 
Specifically, the encoder module identifies Top $T$ optimized path instances, which can approximate the ground truth PathSim, and maps each path instance to an embedding vector. 
The decoder transforms each embedding vector into a scalar respectively, which identifies the similarity score.
We perform extensive experiments on two real-world datasets in different domains, ACM and IMDB. 
Our results demonstrate that NeuPath performs better than state-of-the-art baselines in the PathSim approximation task and similarity search task. 
\end{abstract}

\begin{CCSXML}
<ccs2012>
<concept>
<concept_id>10002950.10003624.10003633.10010917</concept_id>
<concept_desc>Mathematics of computing~Graph algorithms</concept_desc>
<concept_significance>500</concept_significance>
</concept>
<concept>
<concept_id>10002951.10003317.10003338.10003342</concept_id>
<concept_desc>Information systems~Similarity measures</concept_desc>
<concept_significance>500</concept_significance>
</concept>
<concept>
<concept_id>10002951.10003317.10003338.10003346</concept_id>
<concept_desc>Information systems~Top-k retrieval in databases</concept_desc>
<concept_significance>500</concept_significance>
</concept>
</ccs2012>
\end{CCSXML}

\ccsdesc[500]{Mathematics of computing~Graph algorithms}
\ccsdesc[500]{Information systems~Similarity measures}
\ccsdesc[500]{Information systems~Top-k retrieval in databases}

\keywords{Heterogeneous Information Networks, Graph Neural Networks, Similarity Search}

\maketitle

\section{Introduction}
\label{sec-intro}
Heterogeneous information networks (HINs) have been widely used for representing complex real-world applications, such as social networks link prediction \cite{fang2016semantic}, recommendation \cite{yu2014personalized,zhao2017meta,shi2018heterogeneous}, clustering \cite{sun2012integrating,wang2015knowsim}, classification \cite{tang2015pte,wang2016text}, malware detection \cite{hou2017hindroid}, and medical diagnosis \cite{hosseini2018heteromed}, in which multiple typed objects are linked with each other under various relations.  
One of the key principles of the HIN-based mining technique is to use meta-paths, sequences of relations, to guide the processes of similarity search between pairwise nodes in the HIN \cite{sun2011pathsim, dong2017metapath2vec}.  A useful similarity measure should consider both the retrieval accuracy to find most related objects and the generalization ability to handle a growing size of the graph when adding new nodes to it.

\begin{figure}[t]
\centering
\includegraphics[width=0.44\textwidth]{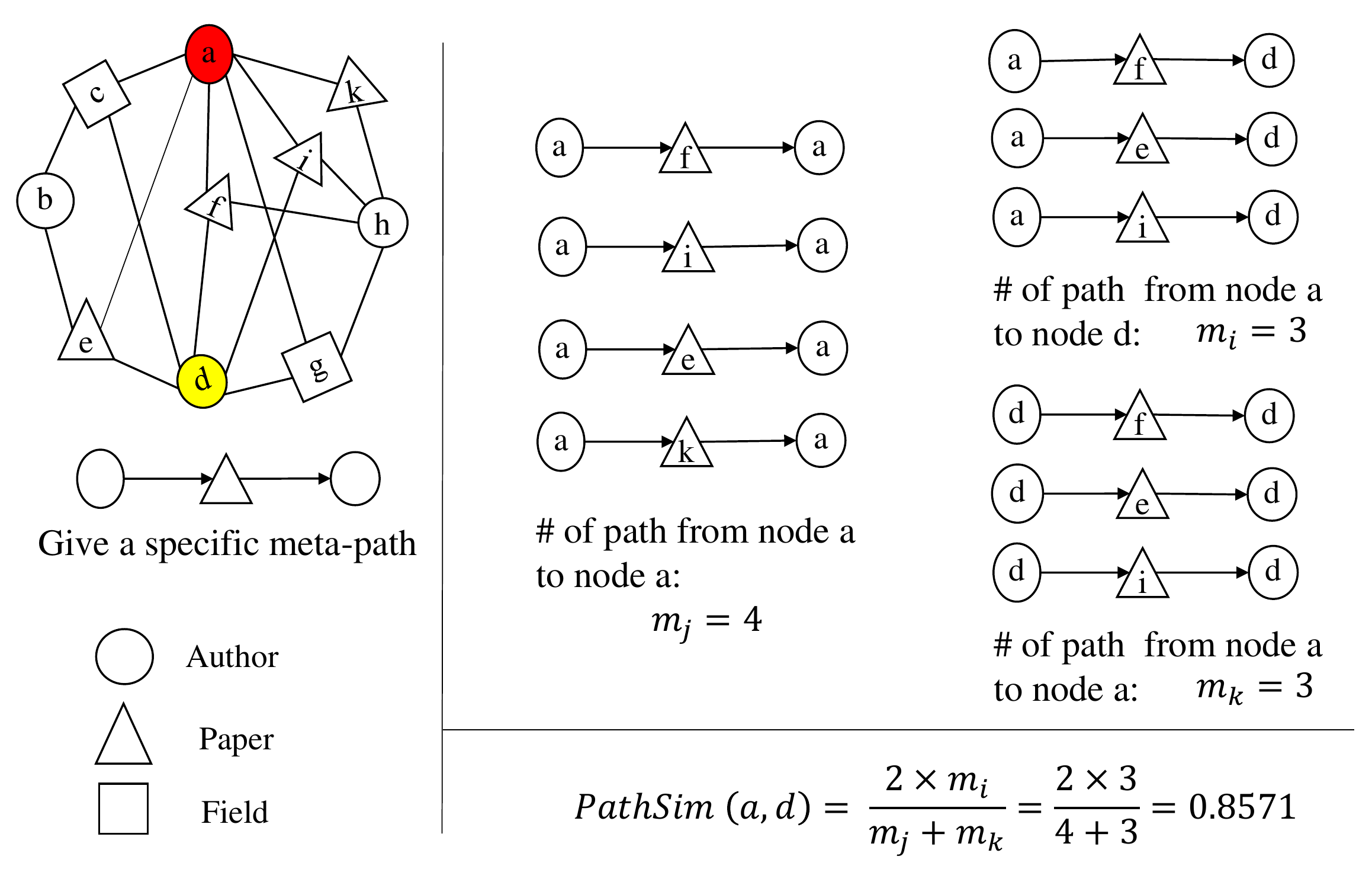}
\caption{An example of calculating the PathSim score between ``a'' and ``d'' under specific meta-path $\gP$. We need to enumerate all the path instances between node ``a'' and ``d'', and count those with meta-path $\gP$. Meanwhile, we do the same process between ``a'' and ``a'', ``b'' and ``b''. Finally, we compute the PathSim(a,d) by Definition \ref{def:pathsim}.} 
\label{fig:example_ps}
\vspace{-0.15in}
\end{figure}

PathSim \cite{sun2011pathsim} and its extensions \cite{meng2015discovering, do2019dw} are considered as fundamental metrics for meta-path-based Top-$K$ similarity search. 
PathSim relies on the number of path instances between nodes following the same meta-path.
When dealing with large HINs, the computation usually involves big matrix multiplication and huge storage of the similarity results.
Moreover, when the network topology changes by adding or deleting nodes or edges, many of the similarity scores should be re-computed.
Thus, an inductive measure is more preferable for meta-path-based Top-$K$ similarity search.

To address this issue, we propose to transform the problem of calculating the PathSim scores into a learning problem. 
A learning based model can summarize its seen examples and has the generalization ability to all other unseen similarities.
In this way, the learning model can be applied to large-scaled HINs and their potential changes.

The straightforward way to use deep learning to approximate PathSim is to make it into a deep learning based method, which uses the ground-truth solutions (PathSim scores) to drive learning, giving the model complete freedom to find a mapping from raw inputs to such solution \cite{xu2020can}. 
Previous works on deep learning for HINs, e.g., HAN \cite{wang2019heterogeneous}, usually consider node classification and link prediction problems on HINs, where they map the objects, regarded as nodes in HIN, into a low-dimensional space where the graph structure and relational information are captured. 
The node representations in these algorithms are relying on local structural information or obtained by recursively aggregating all neighborhood information.
Then the node representations are used as features for calculating the scores by certain similarity functions, e.g., cosine score. 
However, such deep learning algorithms may not be effective to compute PathSim.
As shown in Figure \ref{fig:example_ps}, to compute the PathSim score between node ``a'' and node ``d'' given specific meta-path $\gP$, the main algorithmic process is to first enumerate all the path instances between ``a'' and ``d'', and then select and count the path instances following $\gP$.
Existing deep learning architectures for HINs  based on neighborhood information aggregation ignore the intrinsic algorithmic structure of PathSim, which are based on selected paths, thus failing to find a proper \textit{inductive bias} \cite{battaglia2018relational}\footnote{Inductive bias: a neural network which is more related to the original algorithmic structure can more easily prioritize one solution or interpretation than a neural network that does not.} for PathSim.
Thus, it would be more effective to design a path-based deep model of which the algorithmic structure of PathSim is considered. 

In this paper, to handle the similarity search task, which is to retrieve nodes based on Top-$K$ PathSim scores, we propose Neural PathSim (NeuPath), an encoder-decoder framework to approximate PathSim and automatically inferring the underlying meta-path given a set of node pairs with PathSim scores.  
In the encoding stage, NeuPath identifies a fixed number of path instances that can best infer the same meta-path(s) given in the ground truth PathSim and then maps each path instance to an embedding vector. 
More specifically, the encoder part is designed based on graph neural networks (GNN). 
The features of nodes are denoted as a similarity between the target item and the query item. In each iteration, these embeddings of the features will be updated if the hidden similarity is closer to the ground truth value. In this way, NeuPath tends to search a path instance of which the sequence of relations formulates the same meta-path as in the ground truth PathSim.
In the decoding stage, it leverages these path embeddings and uses a multi-layer perceptron (MLP) to make the embedding be able to approximate the final scores.  

The original PathSim process needs to involve all meta-path instances to compute the final similarity scores. Existing GNN based architectures only record one path instance since at each iteration, it only chooses the neighbor tended to make the most approximate score.  Only one path instance would be not enough to approximate PathSim well while the model might suffer from memory explosion to record all the path instances.  Therefore, we record $T$ path instances with the Top-$T$ closest values to the final score. $T$ is a hyper-parameter for tuning to make feasible results closer to the results by considering all the path instances.  
The parameters are trained in an end-to-end manner, where the model learns the strategy to find the optimized path between two nodes. Therefore, it can effectively generalize the embeddings of similarity on unseen objects by following the learned strategy in NeuPath without training the whole model on the new data again.

The main contributions of this paper are summarized as follows:
\begin{itemize}
	\item We transform the PathSim computation to a learning problem, where an inductive deep learning model is learned to search the most suitable meta-paths in HIN. It releases the restriction of calculating very large graph in traditional PathSim. 
	\item We develop a graph neural network based model, NeuPath. It is a novel approach that represents the path instances rather than the nodes embeddings for the similarity calculation.
	\item We perform extensive experiments on real-world datasets in different domains. Our results demonstrate that NeuPath performs better than state-of-the-art baselines in the similarity search task. Besides, NeuPath can achieve good performance with a small number of training samples.
\end{itemize}

The rest of the paper is organized as follows. We systematically review related work in Section \ref{sec:related_work}. After that, we introduce the architecture of NeuPath in details in Section \ref{sec:neupath}. Section \ref{sec:experiment} presents the evaluation of our model on two real-world datasets.  Finally, we conclude the paper in Section \ref{sec:conclusion}. The notations  in  the  remaining  sections  and  their  descriptions are shown in Table \ref{tb:notation}. The code and data are available at  \href{https://github.com/HKUST-KnowComp/NeuPath}{https://github.com/HKUST-KnowComp/NeuPath}.

\begin{table}[]
\centering
\caption{Notations and Explanations.}
\begin{tabular}{c|c}
\toprule
\textbf{Notation} & \textbf{Explanations}                                                                                                 \\ 
\midrule
$G$               & A Heterogeneous Information Network                                                                                  \\ \hline
$V$               & \begin{tabular}[c]{@{}c@{}}Node set of  $V$, \\ with each node $v$, where $v \in V$\end{tabular}                 \\ \hline

$E$               & \begin{tabular}[c]{@{}c@{}}Edge set of $G$ with each link $e_{uv}$ \\ connecting nodes $u, v$, where $e_{uv} \in E$\end{tabular}            \\ \hline
$T$               & \begin{tabular}[c]{@{}c@{}}The number of path instances identified \\ to represent meta-path in PathSim\end{tabular} \\ \hline
$L$               & \begin{tabular}[c]{@{}c@{}}The number of GNN based iterations \\ with step $l$, where $l \in L$\end{tabular}           \\ \hline
$d$               & \begin{tabular}[c]{@{}c@{}}The dimension of hidden vectors\end{tabular}                 \\ \hline
$\tau (v)$        & Node type for node $v$                                                                                           \\ \hline
$\phi (e)$        & Edge type for edge $e$                                                                                               \\ \hline
$c$               & The query node for similarity search                                                                              \\ \hline
$K$               & The number of most similar nodes  \\
\bottomrule
\end{tabular}
\label{tb:notation}
\vspace{-0.1in}
\end{table}

\section{Related Works}
\label{sec:related_work}
In this section, we briefly summarize the related work in similarity search and graph neural networks for graphs, especially for HINs.

\subsection{Similarity Search in HIN}
% Similarity search is a key task in HIN with various real-world applications like academic network mining \cite{sun2011pathsim,chen2017task}, clustering \cite{sun2012integrating,wang2015knowsim}, classification \cite{tang2015pte,wang2016text}, recommendation \cite{yu2014personalized,zhao2017meta,shi2018heterogeneous}, malware detection \cite{hou2017hindroid}, and medical diagnosis \cite{hosseini2018heteromed}, etc.

Similarity search is an important task in a wide range of real-world applications, such as search engine \cite{page1999pagerank} and recommendation \cite{koren2008factorization}. The key part of similarity search is the measure to compute the similarity of entity pairs.
% Measuring similarities among entities is a significant task in many applications, such as web search \cite{eh2003scaling}, link prediction \cite{lichtenwalter2010new} as well as clustering. 
Similarity search in networks have been explored in the past two decades, which aims to compute node similarity based on the structural information. Representative methods are Personalized PageRank \cite{jeh2003scaling}, SimRank \cite{jeh2002simrank}, SCAN \cite{xu2007scan}, and random walk based ones \cite{tong2006fast,fouss2007random,lao2010fast,lao2010relational,he2019hetespaceywalk}.

Recently, similarity measure focus more on HIN \cite{sun2011pathsim,shi2016survey} since real-world data contains multi-type entities and relations. 
Currently, several works are based on metapaths, which show sequences of node classes and edge types along the paths between two nodes.
Path Count (PC) \cite{sun2011pathsim}, PathSim \cite{sun2011pathsim}, and HeteSim \cite{shi2014hetesim} are some representative meta-path-based measures. However, these methods only suitable for small-scaled graphs or meta-paths with short lengths, while suffering from huge computational challenges when handling large-scale graphs or longer meta-paths.
Some path constraint random walk based models, like Path-Constrained Random Walk (PCRW) \cite{lao2010relational} or metapath2vec \cite{dong2017metapath2vec}, could solve the scalability problem. 
PCRW computes the similarity based on the probability that a walker starting from one node constrained on a particular meta-path reaches the target node.
Metapath2vec develops the meta-path-guided random walk strategy in HIN and then leverages a heterogeneous skip-gram model to perform node embeddings. For similarity search, metapath2vec selects those entities of which embeddings are most similar to the target entity.
However, PCRW and metapath2vec cannot be inductive, i.e., handling unseen nodes, thus failing to be applied to dynamic graphs.

\subsection{Graph Neural Networks}
\begin{figure}[t]
\centering
\includegraphics[width=0.44\textwidth]{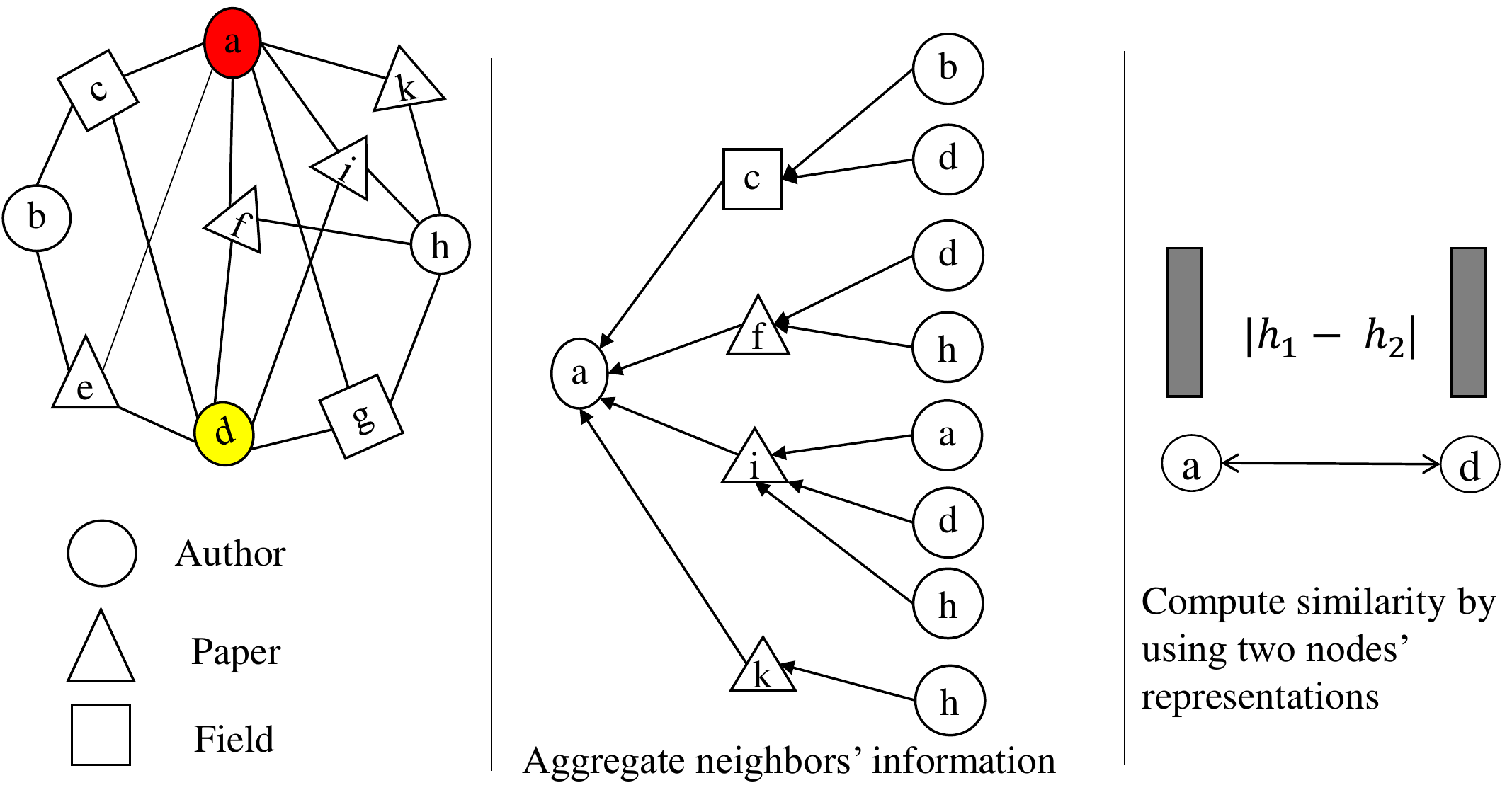}
\caption{The basic process of previous works to learn PathSim. We learn the latent representations of each node first. Then we use similarity function to calculate the scores.} 
\vspace{-0.1in}
\label{fig:node-dependent}
\end{figure}

In recent years, graph neural networks (GNNs) have been state-of-the-art models in graph-based tasks, and representative methods are: GCN \cite{kipf2016semi}, GraphSAGE \cite{hamilton2017inductive}, GAT \cite{velivckovic2017graph}, and GIN \cite{xu2018how}. Most of these works rely on neighborhood aggregation mechanisms \cite{gilmer2017neural}, which is that the representation of a node is learned by iteratively aggregating the embeddings of its neighbors, and thus are inductive to unseen nodes.
Since all of these methods are working with homogeneous graphs, they fail to process HIN, where nodes and edges are of different types, which is more common in reality. Then another line of works extend GNN models to HIN, like RGCN \cite{schlichtkrull2018modeling}, HAN \cite{wang2019heterogeneous}, GTN \cite{Yun2019gtn}, and HGT \cite{hu2020heterogeneous}.
These methods obtain node representations by aggregating neighbors' embeddings implicitly or explicitly defined by meta-path, and then node similarities can be computed by similarity functions like dot production or cosine similarity, the algorithm process shown in Figure \ref{fig:node-dependent}.
While PathSim is a path-based similarity measure, thus, they cannot approximate PathSim in nature.

In \cite{battaglia2018relational}, GNN were shown to provide better inductive bias for relational data, and motivated by this inspiring works, several works propose to use GNN to approximate conventional graph algorithms, like shortest path computation \cite{xu2020can,velivckovic2020Neural,li2020distance}, subgraph counting \cite{liu2020neural} or matching \cite{lou2020neural}, and betweenness centrality computation \cite{fan2019learning}. Based on a learning paradigm, these methods are scalable to large graphs and inductive to unseen nodes. In this work, we propose a novel framework NeuPath to approximate PathSim by considering the computation process of PathSim as an inductive bias to design an effective GNN architecture. To the best of our knowledge, this is the first neural method to learn PathSim \cite{sun2011pathsim}, which addresses the 
aforementioned challenges, i.e., the computational cost and inductive ability, facing conventional methods to compute PathSim in HIN.
Compared to shortest path learning \cite{xu2020can,velivckovic2020Neural}, PathSim requires multiple paths to obtain the similarity, so a mechanism to approximate many path instances should be considered.
Compared to subgraph counting \cite{liu2020neural} or matching \cite{lou2020neural}, our training signal does not contain an explicit pattern (or the meta-path) but asks the learning algorithm to automatically learn the strategy to select nodes along the paths, while for subgraph count or matching, they assume the pattern to count is given.

\section{Methodology: NeuPath}
\label{sec:neupath}
%\subsection{Problem Definition}
In this section, we present the Neural PathSim (NeuPath) to learn the PathSim scores between the query node and other nodes with the same type in the graph. Its idea is to identify several path instances between two nodes by a GNN-based encoder and use these path embeddings to approximate final scores. We first give some definitions in Section \ref{sec:preliminary} and introduce the whole model in Section \ref{sec:problem}, \ref{sec:framework} and \ref{sec:training}. Finally, we present our complexity analysis and the whole algorithm process in Section \ref{sec:analysis}.

\begin{figure*}[t]
\centering
\includegraphics[width=0.95\textwidth]{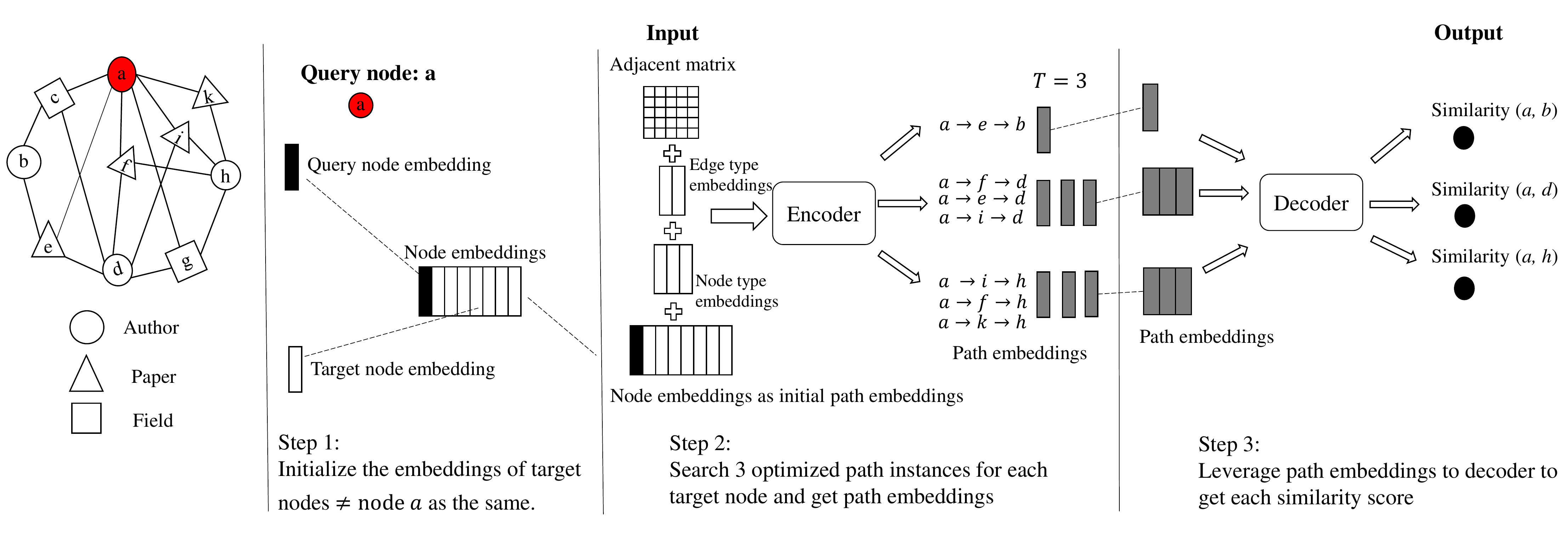}
\caption{The process of NeuPath to get the PathSim similarity for node ``b'', ``d'', ``h'' to the query node ``a''. It contains three steps. At step 1, we construct our node features, which consists of the initial similarity from each node to the query node ``a''. At step 2,  we input the adjacent matrix and the matrix of node features from step 1 to the encoder. Encoder identifies 1 optimized path instances ``a--e--b'' for node ``b'', 3 path instances ``a--f--d'', ``a--e--d'', ``a--i--d'' for node ``d'' and 3 path instances ``a--i--h'', ``a--f--h'', ``a--k--h'' for node ``h''. We map each path instance to an embedding vector. At Step 3, it leverages these path embeddings to two MLP layers to learn PathSim scores for Similarity(a,b), Similarity(a,b), Similarity(a,b) and Similarity(a,h).} 
\label{fig:neupath_process}
\end{figure*}

\subsection{Preliminary}
\label{sec:preliminary}
We first briefly introduce some definitions used in the paper.

\begin{definition}[\textbf{HIN.}]
A \textbf{heterogeneous information network} (HIN) \cite{sun2011pathsim} is a graph $G = (V, E)$ with an node type mapping $\tau$: $V \to \gA$ and a relation type mapping $\phi$: $E \to \gR$, where $V$ denotes the node set, $E$ denotes the edge set, $\gA$ denotes the node type set, and $\gR$ denotes the relation type set. Each node $v \in V$ belongs to one particular node type $\tau(v) \in \gA$, and each edge $e \in E$ belongs to a particular relation $\phi(e) \in \gR$. The number of node types $|\gA|>1$ or the number of relation types $|\gR|>1$.
\end{definition}

\begin{definition}[\textbf{Network Schema.}]
Given a HIN ${G} = (V, E)$ with the node type mapping $\tau$: ${V} \to \gA$ and the relation type mapping $\phi$: $E \to \gR$, the \textbf{network schema} for network $G$, denoted by $\gT_{G} = (\gA, \gR)$, is a graph, in which nodes are node types from $\gA$ and edges are relation types from $\gR$.
\end{definition}

\begin{definition}[\textbf{Meta-path.}]
A meta-path \cite{sun2011pathsim} is a path consisting of a sequence of relations defined between different object types (i.e., structural paths at the meta level). 
A meta-path $\gP$  is a path in the form of $A_{1} \stackrel{R_{1}}{\longrightarrow} A_{2} \stackrel{R_{2}}{\longrightarrow} \cdots \stackrel{{R}_{l}}{\longrightarrow} A_{l+1}$, which describes a composite relation $R = R_{1} \circ R_{2} \circ \cdots \circ R_{l}$ between objects $A_1$ and $A_{l+1}$, where $\circ$ denotes the composition operator on relations.
\end{definition}

\begin{definition}[\textbf{PathSim.}]
\label{def:pathsim}
PathSim is a meta-path-based similarity measure that captures high-order similarity between two objects. Given a symmetric meta-path $\gP$, PathSim between two objects of the same type $x$ and $y$ is:
\begin{equation}
\label{eq:pathsim}
s(x, y)=\frac{2 \times\left|\left\{p_{x \sim y}: p_{x \sim y} \in \mathcal{P}\right\}\right|}{\left|\left\{p_{x \sim x}: p_{x \sim x} \in \mathcal{P}\right\}\right|+\left|\left\{p_{y \sim y}: p_{y \sim y} \in \mathcal{P}\right\}\right|},
\end{equation}
where $p_{x \sim y}$ is a path instance between $x$ and $y$, $p_{x \sim x}$ is that between $x$ and $x$, and $p_{y \sim y}$ is that between $y$ and $y$.
\end{definition}
This shows that given a meta-path $\gP$, $s(x, y)$ is defined in terms of two parts: (1) their connectivity defined by the number of paths between them following $\mathcal{P}$; and (2) the balance of their visibility, where the visibility is defined as the number of path instances between themselves.

\begin{definition}[\textbf{Top-$K$ similarity search under PathSim.}]
Given an information network $G$ and the network schema $\gT_G$, given a meta-path $\gP = \left ( \gP_l \gP_l^{-1} \right )$, where $\gP_l = (A_1 A_2 ... A_l)$, the Top-$K$ similarity search for a node $x_i \in A_1$ is to find sorted $k$ nodes in the same type $A_1$, such that $s(x_i, x_j) \geq  s(x_i, x')$ for any $x'$ not in the returning list and $x_j$ in the returning list, where $s(x_i, x_j)$ is defined as the PathSim score.
\end{definition}

\subsection{Problem Definition}
\label{sec:problem}
Given an information network $G$ and any query node $c$ in type $\tau(c) \in \gA$, the similarity search task is to find the Top-$K$ most similar nodes in type $\tau(c)$ by using the metric of PathSim.
We manually compute the PathSim score $b_{ct}$ between node $c$ and $t$, where $\tau(t) = \tau(c)$ by  Equation (\ref{eq:pathsim}) under a meta-path $\gP$. The goal is to learn the strategy of computing similarity score $y_{ct}$ such that $y_{ct} = b_{ct}$.

In this way, we only need to calculate the PathSim scores as the ground truth data for a small number of nodes in the graph and use these samples to train NeuPath and learn the strategy of computing PathSim. Then the trained NeuPath can automatically compute PathSim scores for the rest of nodes in the graph in parallel. 
% Retrieve some nodes' PathSim. Get others to learning strategies

\subsection{Basic Framework}
\label{sec:framework}
We consider an HIN $G = (V, E)$ where $V$ is the set of nodes and $E$ is the set of edges, where each node $v \in V $ has associated node feature $x_v$ and node type feature $\tau(v)$.
Similarly, each edge $e \in E $ has associated edge type features $\phi(e)$. 

% We use [dv, 1, 1] as node v’s initial feature $Xv$, and let $h(0)$
% v = Xv. The neighborhood of node v, N(v), is defined as all the nodes that are adjacent to v, and h(l)
% N(v) denotes the aggregated
% neighborhood representation output by the l−th layer of the model

Give a query node $c$, we set the initial node feature embedding $\vx_t \in \sR^d$ of node $t$ with the same node type shown as follows: 
\begin{equation} 
\vx_t=
\left\{\begin{array}{ll}
\left [ 1,0,0,...,0 \right ] & t=c, \\
\left [ 0,1,0,...,0 \right ] & t \neq c.
\end{array}\right.
\end{equation}
We set all the initial vectors for nodes as the same except for the query node $c$ itself. 

We use $\vh_t^l[i] \in \sR^d$ to denote the $i$-th embedding of node $t$ at the $l$-th layer of the Encoder, regarding as the path embedding for the PathSim score between target node $t$ and query node $c$ in $l$-th iteration of Encoder. 
Here, $d$ is the dimension of hidden embeddings, which we assume to be the same across different layers for the sake of simplicity, $l \in \gL,  \gL = \left \{ 1, 2, ..., L \right \}$ and 
$i \in \gT, \gT = \left \{ 1, 2, ..., T \right \}$ . Let $\vh_t^{0}[i] = \vx_t, \; \mathrm{where} \; \forall i \in \gT$.  

Then, we input node features, node type information, edge type information, graph adjacent matrix into an Encoder introduced in Section \ref{sec:encoder}. The goal of it is to search $T$ path instances that could best represent the meta-path $\gP$ for each node $i$, respectively, 
Finally, the decoder module is designed as a multi-layer perceptron (MLP) to make the embedding be able to approximate the final scores, introduced in Section \ref{sec:decoder}.

We give a concrete example of NeuPath for approximating PathSim scores among the query node ``a'' and node ``b'', ``d'', ``h'' with the same node type in Figure \ref{fig:neupath_process}.

\subsection{Encoder}
\label{sec:encoder}
\begin{figure}[t]
\centering
\includegraphics[width=0.44\textwidth]{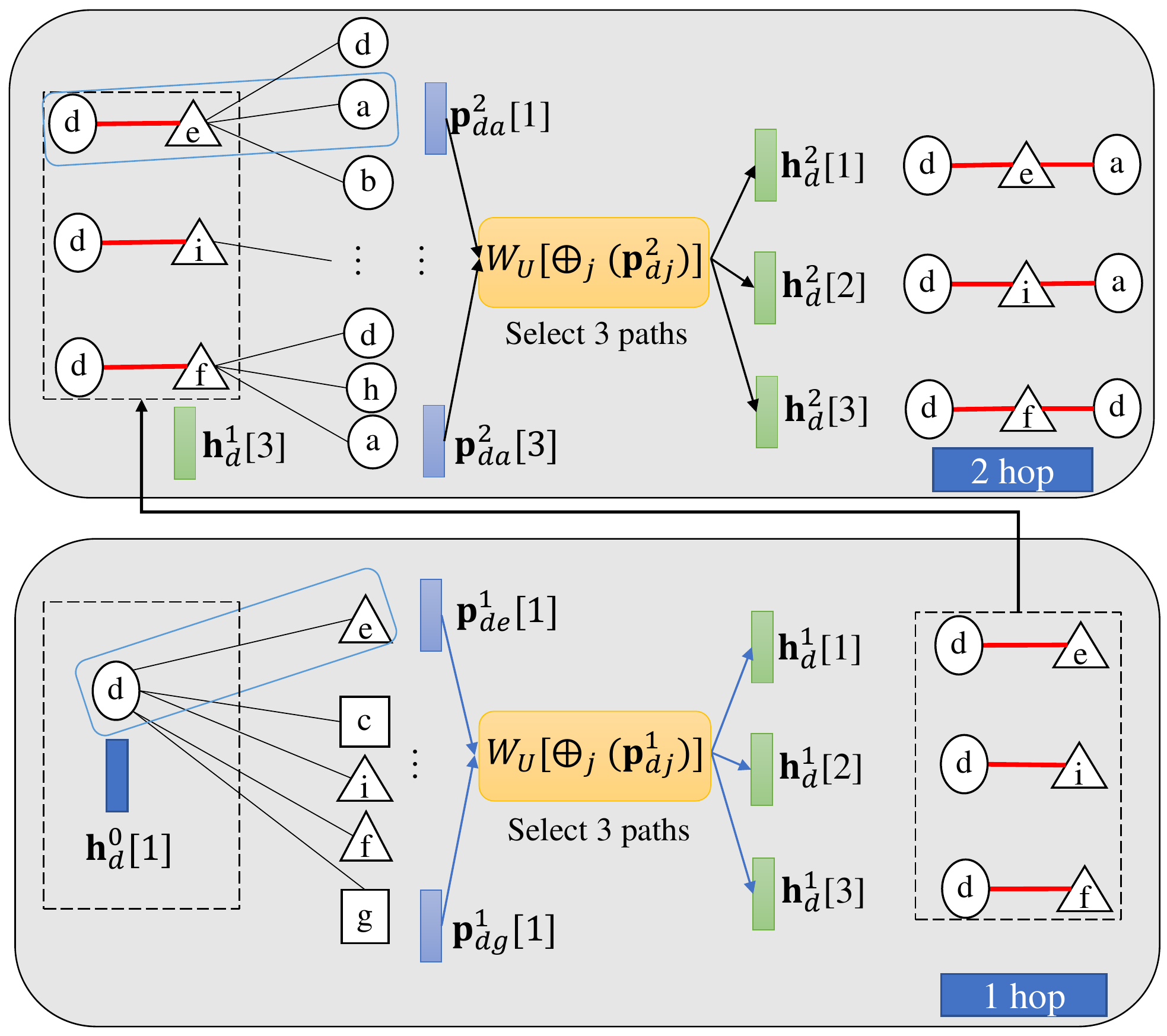}
\caption{The process of Multi-Path Encoder to identify 3 path instance between node ``a'' and ``d''. By first iteration of GNN layer, the encoder generates 3 path instances with length of 2, shown as ``d--e'', ``d--i'', ``d--f''. By second iteration, the encoder generates 3 path instances with length of 3, shown as ``d--e--a'', ``d--i--a'', ``d--f--a''.} 
\label{fig:encoder}
\vspace{-0.1in}
\end{figure}

As is indicated in Figure \ref{fig:example_ps}, computing a node’s PathSim score needs to iteratively aggregate its neighbors' information, which is similar to the neighbor aggregation in graph neural networks. 
For the specific design, we consider three components: \textbf{Extract}, \textbf{Compare}, and \textbf{Update}. We introduce them in details in the following parts. Besides, we give an example of search $T=3$ path instances between node $a$ and node $d$ by a 2-layer encoder, shown in Figure \ref{fig:encoder}.

\subsubsection{Extract}
We extract the message by using the neighbor node $s$ and the edge $e_{st}$. We first design the node type-specific transformation matrix $\mW_\tau \in \sR^{d \times d}$ to project the features of different types of nodes into the same feature space. The projection process can be shown as follows:
\begin{equation}
\begin{aligned}
\tilde{\vh}_s^{(l-1)}[i] &=\mW_{\tau(s)}\left(\vh_s^{(l-1)}[i]\right), \\
\tilde{\vh}_t^{(l-1)}[i] &=\mW_{\tau(t)}\left(\vh_t^{(l-1)}[i]\right).
\end{aligned}
\label{eq:extract_1}
\end{equation}

Then, information is passed from node $s$ to $t$ through edge $e_{st}$ by the following equation:
\begin{equation}
    \vp_s^{(l)}[i] = \mW_m \left( \tilde{\vh}_s^{(l-1)}[i] \mathbin\Vert \ve_{\phi(e)} \mathbin\Vert \tilde{\vh}_t^{(l-1)}[i]\right),
\label{eq:extract_2}
\end{equation}
where $\ve_{\phi(e)}$ is a trainable vector to represent edge type $\phi(e)$, $\mW_m \in \sR^{d \times 3d}$ is a matrix. $||$ is an operator for concatenation.

\subsubsection{Compare}
We compare the values of $(T \times |\gN(t)|)$ path instances and select $T$ path instances with Top-$T$ closest scores to the ground truth PathSim score. The equations is as follows:
\begin{eqnarray}
\vq_t^{(l)}[i] = \bigoplus_{(s, t) \in E, i \in \gT} \vp_s^{(l)}[i],
\label{eq:compare}
\end{eqnarray}
where $\oplus$ is an element-wise aggregation operator. To get $T$ path instances, we use $T$-max pooling as $\oplus$. 

\subsubsection{Update}
We compare the values of $T$ path instance embeddings with the three previous values at $l-1$ step and update the path embeddings.
\begin{equation}
%&&\vec{h}_{i}^{(t)}=\operatorname{ReLU}\left(\sum_{j, i) \in E} a\left(\vec{z}_{i}^{(t)}, \vec{z}_{j}^{(t)}, \vec{e}_{i j}^{(t)}\right) \mathbf{W} \vec{z}_{j}^{(t)}\right) \quad \\
\vh_t^{(l)}[i] = \mW_u \left(\vh_t^{(l-1)}[i] \mathbin\Vert \vq_t^{(l)}[i]\right),
\label{eq:update}
\end{equation}
where $\mW_u \in \sR^{d \times 2d}$ is a trainable matrix used to combine the message from the last layer and this layer.

\subsection{Decoder}
\label{sec:decoder}
The decoder is implemented with a two layered MLP which maps the embedding $\vh_t^{(l)}[i]$ to the approximate PathSim score $y_{ct}$:
\begin{equation}
y_{ct} = \mW_1 \left(RELU \mW_2 \left( \mathbin\Vert_{i \in \gT} \vh_t^{(L)}[i] \right)\right),
\label{eq:decode}
\end{equation}
where $\mW_2 \in \sR^{d \times dT}$ is a weight trainable matrix, $\mW_1 \in \sR^{1 \times d}$ is a trainable vector to predict final score, $RELU$ is the activation function.

\subsection{Training Algorithm}
\label{sec:training}
There are two sets of parameters to be learned, including those in encoder and decoder. We use the following mean squared error (MSE) loss to update these parameters. Given a node pair $(c, t)$, suppose the ground truth PathSim value is $b_{ct}$, our model learn to infer $y_{ct}$. The training loss is defined as: 
% the squared error $S_{ct}$ between $b_{ct}$ and $y_{ct}$ is
% \begin{equation}
% S_{c, t}= (y_{ct} - b_{ct})^2.
% \end{equation}
\begin{equation}
\text { Loss }= \frac{1}{M}\sum_{t \in V} (y_{ct} - b_{ct})^2,
\end{equation}
where $M$ is the number of training pairs for the query node $c$. 
%$M = C \times K$, where $C$ is the number of query node $c$ and $K$ is the number of most similar nodes in ranking lists.

With the guide of labeled data, we can optimize the proposed model via backpropagation and learn the embeddings of nodes.

\subsection{Complexity Analysis and Algorithm}
\label{sec:analysis}
During training, the time complexity is proportional to the training iterations $L$ and the number of query nodes. Since NeuPath is trained to learn the strategy of identifying proper path instances which are helpful to approximate the final PathSim scores, the training phase is performed only once here, and we can utilize the trained model for any input network with a certain query node in the application phase. 

In the similarity search task to infer the ranking results from the trained model, the time complexity consists of two parts. The first is from the encoding phase. The encoder complexity takes $O(L|V|N(\cdot))$, where $L$ is the number of iterations, $V$ is the number of nodes, $N(\cdot)$ is the average number of node neighbors. The second is from the decoder and top-$K$ values' retrieval. Once the nodes are encoded, we can compute their respective PathSim score, and return the top-$k$ highest nodes. The time complexity of this process mainly comes from the sorting operation, which takes $O(|V|\text{log}(|V|))$. Therefore, the total time complexity for the similarity search should be $O(L|V|N(\cdot)+|V|\text{log}(|V|)$.

We give the algorithm on how NeuPath process to make approximation in Algorithm \ref{alg:neupath}.

\begin{algorithm}[!t]
\caption{NeuPath.}
%\setstretch{1.2}
\begin{algorithmic}[1]
        \REQUIRE {A graph $G = (V, E)$, query node $c$, node features $\left\{\vx_{v}, \forall v \in V\right\}$}
        \ENSURE PathSim scores $y_{ct}$ for each node $t$ with the same node type as query node $c$

\STATE {/* Encoding Path Embeddings */}
\FOR{$i = 1$ to $T$}
\STATE{$\vh_v^{(l)}[i] \leftarrow \mathbf{x}_{v}$}
\ENDFOR
\FOR{$l = 1$ to $L$}
    \FOR{$v \in V$}
	\STATE{Update $\vh_v^{(l)}[i]$ according to Eqs. (\ref{eq:extract_1}), (\ref{eq:extract_2}), (\ref{eq:compare}), (\ref{eq:update}})
	\ENDFOR
\ENDFOR
\STATE {/* Decoding for PathSim scores */}
    \FOR{$t \in V$ where $\tau(t) = \tau(c)$}
    \STATE{Get Pathsim scores $y_{ct}$ according to Eq. (\ref{eq:decode}})
    \ENDFOR
    
\end{algorithmic}
\label{alg:neupath}
\end{algorithm}

\section{Experiment}
\label{sec:experiment}
In the section, we conduct experiments to answer the following questions:
\begin{itemize}
	\item Q1: Given PathSim based ranking lists of some nodes under a certain meta-path $\gP$, whether NeuPath can get appropriate ranking lists of other nodes?
% 	\item Q2: Whether the model can inter the semantic rules given the proximities in the specific graph?
    \item Q2: How do the components of NeuPath, such as aggregating operator, and different parameter values affect its estimation?
	\item Q3: Whether NeuPath has generalization ability?
\end{itemize}

We give the experimental settings including evaluation metrics, dataset description, baselines description as well as the implementation details of NeuPath in Section \ref{sec:metric}, \ref{sec:data_des}, \ref{sec:baselines} and \ref{sec:implementation}, respectively. We conduct the experiment on PathSim approximation and similarity search, mentioned in Section \ref{sec:pathsim_approximate}, \ref{sec:sim_search} to answer Q1. In addition, we do ablation study in Section \ref{sec:ablation} and \ref{sec:hyper} to answer Q2. Finally, we answer Q3 by analyzing the influence of the scale of training samples on approximating effectiveness, indicated in Section \ref{sec:running_time} and Section \ref{sec:generalization}.   

\subsection{Evaluation Metric}
\label{sec:metric}
\subsubsection{\bf{RMSE for PathSim Approximation.}}
We choose Root Mean Square Error (RMSE) to evaluate the quality of NeuPath's approximating the ground truth PathSim scores since it is a frequently used measure of the differences between values predicted by a model and the values observed. It is defined as:
\begin{equation}
\text{RMSE} = \sqrt{\frac{1}{|\bar{\Omega}|} \sum_{(i,j) \in \bar{\Omega}}{(\tO_{ij} - \tR_{ij})^2}},
\end{equation} 
where $\bar{\Omega}$ is the set of all user-item pairs $(i,j)$ in the test set, $\tO_{ij}$ is the value observed,  
and $\tR_{ij}$ is the corresponding value predicted by the model.

\subsubsection{\bf{nDCG for Similarity Search.}}
we use the measure nDCG (Normalized Discounted Cumulative Gain) \cite{jarvelin2002cumulated} to evaluate the power of similarity search by comparing its output ranking results with the labeled ones.  Given a list of nodes ranked by predicted scores, and their graded relevance values, discounted cumulative gain at position $k$ (DCG@$k$) is defined as:
\begin{equation}
\operatorname{DCG} @ k=\sum_{i=1}^{k} \frac{r_{i}}{\log _{2}(i+1)},
\end{equation}
where $r_i$ denotes the graded relevance of the node at position $i$. 
%Due to the logarithmic reduction factor, the gain $r_i$ of each node is penalized at lower ranks.
Consider an ideal DCG at rank position $k$ (IDCG@$k$) which is obtained by an ideal ordering of nodes based on their relevance scores. Normalized DCG at position
$k$ (nDCG@$k$) is then computed as:
\begin{equation}
\operatorname{nDCG} @ k=\frac{\operatorname{DCG} @ k}{\operatorname{IDCG} @ k}.
\end{equation}
Our motivation for using nDCG@$k$ is to test the quality of ranking for the top $k$ entities.
In this paper, we set $k=20$.

For RMSE, the lower the better, and for nDCG, the higher the better.
%The authors are divided into four areas: database, data mining, machine learning, information retrieval. Also, we label each author’s research area according to the conferences they submitted.

\subsection{Dataset Description}
\label{sec:data_des}
In this section, we present the statistics of the two datasets (ACM and IMDB) and show how to gain the samples for training, validation, and testing.

\subsubsection{\bf{Statistics of dataset.}}
For ACM dataset, it contains 31,847 nodes from four node types, including papers (P), authors, conferences (C), and fields (F).  Also, it has 79,987 edges from three relations/edge types, including Author-Paper (A-P), Author-Field (A-F), and Paper-Conference (P-C). 
For IMDB dataset, it contains 12,890 nodes from four node types, including papers (P), authors, conferences (C), and fields (F).  Besides, it has 19,120 edges from two relations/edge types, including Movie-Actor (M-A), Movie-Director (M-D). 
The detailed sizes of each item and each relation are shown in Table \ref{tb:statistics}.

\subsubsection{\bf{Ground truth discussion.}}
We build PathSim scores as the ground truth between each query node and other nodes to evaluate models' power of PathSim approximation and similarity search, respectively.  For both two datasets, the sizes of the training nodes, validation nodes and the testing nodes are 400, 100 ,400 respectively for each meta-path. The number of triplets for training pairwise similarity under is 4000 under each meta-path setting. The experiments in the paper compare the power of predicting a large number of PathSim scores with only manually constructing a small number of the training data.

For ACM, the number of triplets for testing paper similarity is 4,999,600 (the number of testing node multiples the number of paper in HIN). Also, the number of triplets for testing author similarity is 6,972,400.
For IMDB, the number of triplets for testing movie similarity is 1,912,000. Also, the number of triplets for testing director similarity is 907,600.

\begin{table}[t]
\centering
\caption{Statistics of dataset ACM and IMDB. The table presents the size of each item and relation in two datasets.}
\resizebox{0.45\textwidth}{!}{
\begin{tabular}{c|cccc}
\toprule
\multirow{4}{*}{ACM}  & Author (A) & Paper (P) & Conference (C) & Field (F) \\
                      & 17,431     & 12,499    & 14             & 1,903     \\
                      \cline{2-5}
                      & A-P        & A-F      & P-C           &           \\
                      & 37,055     & 30,424    & 12,499         &           \\
                      \midrule
\multirow{4}{*}{IMDB} & Movie (M)  & Actor (A) & Director (D)   &           \\
                      & 4,780      & 5,841     & 2,269          &           \\
                      \cline{2-5}
                      & M-A        & M-D       &                &           \\
                      & 14,340     & 4,780     &                &          \\
\bottomrule
\end{tabular}
}
\label{tb:statistics}
\end{table}

\begin{table}[t]
\caption{Different meta-paths constructed by expert knowledge for two Datasets, applied to HAN. Since PathSim is confined merely on the symmetric meta-paths, we just construct round trip meta-paths.}
\resizebox{0.45\textwidth}{!}{
\begin{tabular}{c|c|c}
\toprule
 & \textbf{Candidate} & \textbf{Ground Truth} \\ 
 \midrule
\multirow{2}{*}{\textbf{ACM}} & APA, AFA, AFAFA, APCPA, APAPA  & APA, APCPA \\ \cline{2-3} 
 & PCP, PAP, PCPCP, PAFAP, PAPAP & PCP, PAFAP \\ \hline
\multirow{2}{*}{\textbf{IMDB}} & MDM, MAM, MDMDM, MAMAM & MAM, MAMAM \\ \cline{2-3} 
 & DMD, DMDMD, DMAMD & DMD, DMAMD \\ 
 \bottomrule
\end{tabular}
}
\vspace{-0.1in}
\label{tb:HAN}
\end{table}

\begin{table*}[t]
\caption{Comparison of different methods on approximating PathSim scores. We use RMSE for evaluation, the lower, the better. 
% For meta-path ``DMD'' since the node would link to itself with meta-path ``DMD'', it is meaningless to compute the similarity between directors under this meta-path.
}
\resizebox{0.75\textwidth}{!}{%
\begin{tabular}{c|cccc|cccc}
\toprule
\multirow{2}{*}{Models} & \multicolumn{4}{c|}{ACM}           & \multicolumn{4}{c}{IMDB}  \\
\cline{2-9}
                        & APA    & APCPA  & PCP    & PAFAP  & MAM & MAMAM & DMD & DMAMD \\
\midrule
MLP                     & 0.2835 & 0.0439 & 0.0314 & 0.2408 &  0.3790   &   0.3354    &  0.1708  &    0.2949   \\
GAT                     & 0.2519 & 0.0381 & 0.0963 & 0.2378 &  0.4023   &   0.3567    &  0.1121   &  0.2991     \\
GCN                     & 0.2400 & 0.0374 & 0.0818 & 0.2036 &   0.3887  &  0.3320     &  0.1266   &  0.3072     \\
HAN                     & 0.3301 & 0.0321 & 0.0012 & 0.3091 &  0.4444   &   0.4167    &  0.1236   &  0.2881     \\
RGCN                    & 0.2289 & 0.0319 & 0.0042 & 0.3267 &   0.3844  &   0.3676    &  0.1273   &  0.3065     \\
HGT                     & 0.3428 & 0.0311 & \textbf{0.0003} & 0.3354 &  0.4561   &  0.3919     &  0.1247 &   0.3018    \\
GTN                     & 0.2232 & 0.0329 & 0.0007 & 0.3219 &  0.3236   &   0.3456    & 0.1257    &  0.3321     \\
\hline
NeuPath                 & \textbf{0.1902} & \textbf{0.0210}  & 0.0005 & \textbf{0.1955} &  \textbf{0.3001}   &   \textbf{0.3111}    & \textbf{0.1101}    &  \textbf{0.2506}     \\
\bottomrule
\end{tabular}
}
\label{tb:rmse}
\end{table*}

\begin{table*}[t]
\caption{Comparison of different methods on similarity search task. We use nDCG for evaluation, the higher, the better. 
% For meta-path ``DMD'' since the node would link to itself with meta-path ``DMD'', it is meaningless to compute the similarity between directors under this meta-path.
}
\resizebox{0.75\textwidth}{!}{%
\begin{tabular}{c|cccc|cccc}
\toprule
\multirow{2}{*}{Models} & \multicolumn{4}{c|}{ACM}           & \multicolumn{4}{c}{IMDB}  \\
\cline{2-9}
                        & APA    & APCPA  & PCP    & PAFAP  & MAM & MAMAM & DMD & DMAMD \\
\midrule
MLP                     & 0.5472 & 0.1665 & 0.1657 & 0.1969 &  0.5497   &   0.3462    &   0.1690  &   0.5471    \\
GAT                     & 0.4905 & 0.1700 & 0.1721 & 0.2117 &  0.4133   &   0.3132    &   0.1779  &   0.5335    \\
GCN                     & 0.5150 & 0.1725 & 0.1716 & 0.2208 &  0.4546   &   0.3486    &   0.1725  &  0.5321     \\
RGCN                    & 0.4717 & 0.1737 & 0.1699 & 0.2149 &  0.4265   &   0.3287   &   0.1722  &   0.5498    \\
HAN                     & 0.5777 & 0.1605 & 0.1923 & 0.2232 &  0.5329   &   0.3932   &   0.1770  &   0.5674    \\
HGT                     & 0.5632 & 0.1663 & 0.1816 & 0.2151 &  0.5467   &   0.3433    &   0.1791  &   0.5358    \\
GTN                     & 0.5812 & 0.1722 & 0.2001 & 0.2341 &  0.5605   &    0.4032   &   0.1801  &    0.5534   \\
\hline
NeuPath                 & \textbf{0.6115} & \textbf{0.2301} & \textbf{0.2441} & \textbf{0.2405} &  \textbf{0.5832}   &    \textbf{0.4309}   &  \textbf{0.1821}   &  \textbf{0.5998}    \\
\bottomrule
\end{tabular}
}
\label{tb:ndcg}
\end{table*}

\subsection{Baselines}
\label{sec:baselines}
To evaluate the approximation performance of NeuPath, we compare it with three kinds of deep learning based methods: neural network based method ignoring the graph structure (MLP), GNN based models on homogeneous graph (GCN, GAT), and HIN based methods utilizing all the heterogeneous information (RGCN, HAN, HGT, GTN). We ignore those unsupervised graph-related methods such as node2vec \cite{grover2016node2vec} and metapath2vec \cite{dong2017metapath2vec} because there are supervised signals (ground truth data) in the experiments. All these methods learn the node representations and then calculate PathSim scores by similarity functions. Since we only have the graph structure in the experiments, the original node features and edge features used by baselines are initialized randomly. 

The details of baselines are as follows.

\textbf{MLP}\cite{santoro2017simple} This function follows a general paradigm: simply concatenate the node feature embeddings as well as the related edge features embeddings together and feed them into MLP layers.

\textbf{GCN \cite{kipf2016semi}.}
It is a semi-supervised graph convolutional network designed for homogeneous graphs. Since GCN does not consider the exact relations, we only input the adjacent matrix and node features matrix for training while ignoring the node types and edge types.

\textbf{GAT \cite{velivckovic2017graph}. }
It is a semi-supervised neural network that considers the attention mechanism on homogeneous graphs. The inputs for training are similar to GCN, we just input the adjacent matrix and node features matrix for training and do not use the node types and edge types.

\textbf{HAN \cite{wang2019heterogeneous}.}
It is a meta-path based heterogeneous graph neural network based on hierarchical attention. Since HAN needs expert knowledge to design some candidate meta-paths to make the heterogeneous graph into several homogeneous graphs, we enumerate all the possible meta-paths for training. The candidate meta-paths are shown in Table \ref{tb:HAN}.

\textbf{RGCN \cite{schlichtkrull2018modeling}. }
It sets a different weight for each relationship, i.e., a relation triplet. We apply the original RGCN which uses an adjacent matrix, node features matrix as well as relations between nodes (edge types). We use the implementation provided in DGL.

\textbf{HGT \cite{hu2020heterogeneous}. }
It uses the meta relations of heterogeneous graphs to parameterize weight matrices for the heterogeneous mutual attention, message passing, and propagation steps.

\textbf{GTN \cite{Yun2019gtn}. }
Graph Transformer Layer (GTL) in GTN softly selects adjacency matrices (edge types) from the set of adjacency matrices and generates a new meta-path graph via the matrix multiplication of two selected adjacency matrices.

\textbf{NeuPath.}
 If without any clarification, it is the model using $T$-max pooling operator and all node types, edge types features.

\subsection{Implementation Details}
\label{sec:implementation}
We use $d=256$ as the hidden dimension throughout the neural networks for all baselines. For the NeuPath Encoder, we set the path instance number $T$ as 3 for the training samples with meta-path ``APCPA'', ``PAFAP'', and set $T=2$ for other training. We try $L=2$ and $L=4$ layers of GNN based Encoder for training NeuPath, since the meta-path in PathSim is defined to be symmetric.  All baselines are optimized via the AdamW optimizer with the Cosine Annealing Learning Rate Scheduler. For NeuPath, we train it for 10 epochs and select the one with the lowest validation loss as the reported model.

\vspace{-0.1in}

\subsection{PathSim Approximation}
\label{sec:pathsim_approximate}
In this section, we evaluate the power of NeuPath to approximate the ground truth PathSim under different meta-paths. 

From Table \ref{tb:rmse}, we compare the exact values predicted by baselines with the ground truth PathSim scores by using RMSE results. It shows that NeuPath could get the highest RMSE results compared to other baselines except for meta-path ``PCP''. It proves that a model of which architecture is more related to the PathSim process can better approximate the PathSim scores. Moreover, the reason that HGT achieves the lowest RMSE on meta-path "PCP" might be that HGT is the state-of-art GNN based model for learning node representation in HIN and it can incorporate information from high-order neighbors of different types through message passing across layers, which can be regarded as “soft” meta-paths. Therefore, it also searches appropriate meta-path similar to the ground truth meta-paths.
Besides, there is a less obvious difference on RMSE among the simplest model (MLP) and complex GNN based model (HGT), which indicates a model could not improve the performance a lot even though the model has a more complex architecture.

\subsection{Similarity Search}
\label{sec:sim_search}
Since PathSim is a commonly used measure for similarity search, in this section, we evaluate the effectiveness of different models to conduct the similarity search task. 

Table \ref{tb:ndcg} shows that NeuPath gains the highest nDCG@$10$ results compared to other baselines. Moreover, we find that the three HIN-based methods achieve comparable performances, all worse than NeuPath. It indicates that even though HIN-based methods can capture the structural information for nodes in the graph, the models still could not prioritize others significantly if their architectures do not consider the original process of PathSim. In other words, all these GNN based methods (GAT, GCN, RGCN, HAN, HGT, GTN) do not find a proper inductive bias to learning PathSim. Therefore, they could not consistently outperform MLP, the basic neural network.

For meta-path ``DMD'' since the node would link to itself with meta-path ``DMD'', it is meaningless to compute the similarity between directors under this meta-path. RMSE and nDCG results on meta-path ``DMD'' are relatively smaller than others since only two scores (1 for similarity to query node itself, 0 for others) are used as ground truth PathSim scores.

\subsection{Ablation Study}
\label{sec:ablation}
We further compare a variant of our NeuPath model in terms of the following aspects to demonstrate the effectiveness of the framework design: the use of $T$-max pooling operator, the use of node type features, and the use of edge type features. The results shown in Table \ref{tb:ablation_approximation} and Table \ref{tb:ablation_similarity}. We can conclude as follows.

\begin{itemize}
    \item Without using $T$-max pooling operator to identify $T$ path instances, the operator can be replaced by mean-pooling, max-pooling, or sum-pooling, outputting only one path instance. This will increase RMSE scores and decrease nDCG consistently on four meta-path based data, which indicates the efficacy to use more than one path to approximate PathSim.
    
    \item The ignoring considering node type features also make the performance worse than NeuPath, which proves the importance of projecting nodes into the same feature space.
    
    \item We test how edge type affects the performance. The influence of edge type on the final results is larger than others. The reason is that the model needs to use edge type (edge information) to construct path instances.   
\end{itemize}

\begin{table}[t]
\centering
\caption{Performance of PathSim approximation by different variants of NeuPath on ACM dataset. We give RMSE results.}
\resizebox{0.45\textwidth}{!}{%
\begin{tabular}{c|cccc}
\toprule
Variants                   & APA             & APCPA           & PCP             & PAFAP           \\ 
\midrule
mean-pooling               & 0.2013          & 0.0274          & 0.0006          & 0.2104          \\
max-pooling                & 0.1991          & 0.0300          & 0.0005          & 0.2023          \\
sum-pooling                & 0.1956          & 0.0288          & 0.0005          & 0.2066          \\ \hline
w/o node type              & 0.2001          & 0.0212          & 0.0005          & 0.2027          \\
w/o edge type              & 0.2239          & 0.0475          & 0.0007          & 0.2299          \\
w/o node \& edge type & 0.2247          & 0.0490          & 0.0007          & 0.2302          \\ \hline
NeuPath                    & \textbf{0.1902} & \textbf{0.0210} & \textbf{0.0005} & \textbf{0.1955} \\ 
\bottomrule
\end{tabular}
}
\label{tb:ablation_approximation}
\end{table}

\begin{table}[t]
\centering
\caption{Performance of similarity search by different variants of NeuPath on ACM dataset. We give nDCG results.}
\resizebox{0.45\textwidth}{!}{%
\begin{tabular}{c|cccc}
\toprule
Variants                   & APA    & APCPA  &  PCP    & PAFAP  \\
\midrule
mean-pooling               & 0.5901 & 0.2009 & 0.2089 & 0.2016 \\
max-pooling                & 0.5888 & 0.2176 & 0.2227 & 0.2224 \\
sum-pooling                & 0.5833 & 0.2143 & 0.2199 & 0.2126 \\
\hline
w/o node type              & 0.6002 & 0.2121 & 0.2339 & 0.2327 \\
w/o edge type              & 0.5306 & 0.1764 & 0.1887 & 0.2249 \\
w/o node \& edge type & 0.5106 & 0.1735 & 0.1701 & 0.2148 \\
\hline
NeuPath                    & \textbf{0.6115} & \textbf{0.2301} & \textbf{0.2441} & \textbf{0.2405} \\
\bottomrule
\end{tabular}
}
\vspace{-0.1in}
\label{tb:ablation_similarity}
\end{table}

\vspace{-5pt}
\subsection{Hyper-parameter Sensitivity}
\label{sec:hyper}
\begin{figure}[t]
    \centering
    \subfigure[\# of path instance $T$ under ``APA''.]{\includegraphics[width=0.23\textwidth, height=0.14\textheight]{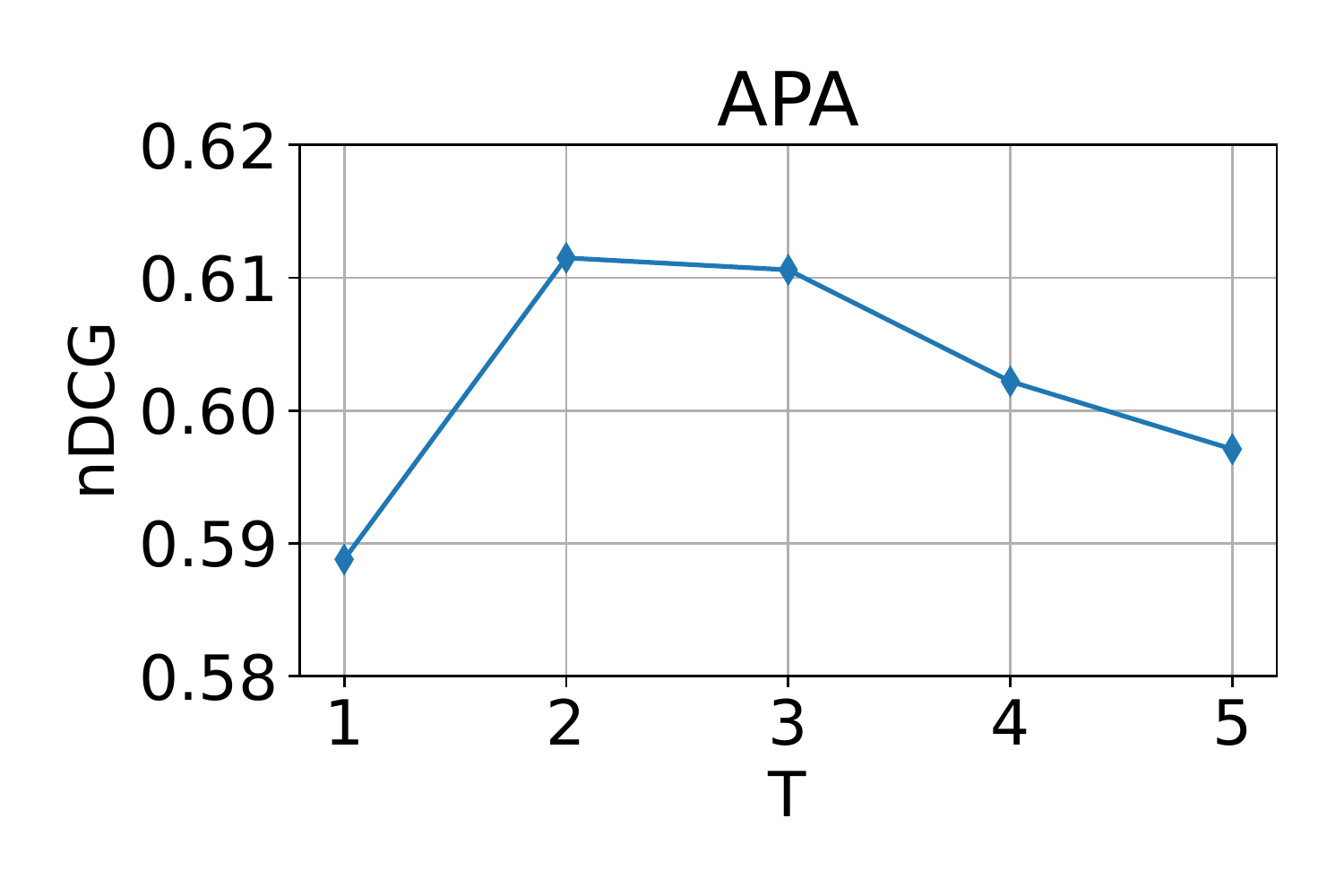}}
    \subfigure[\# of path instance $T$ under ``APCPA''.]{\includegraphics[width=0.23\textwidth, height=0.14\textheight]{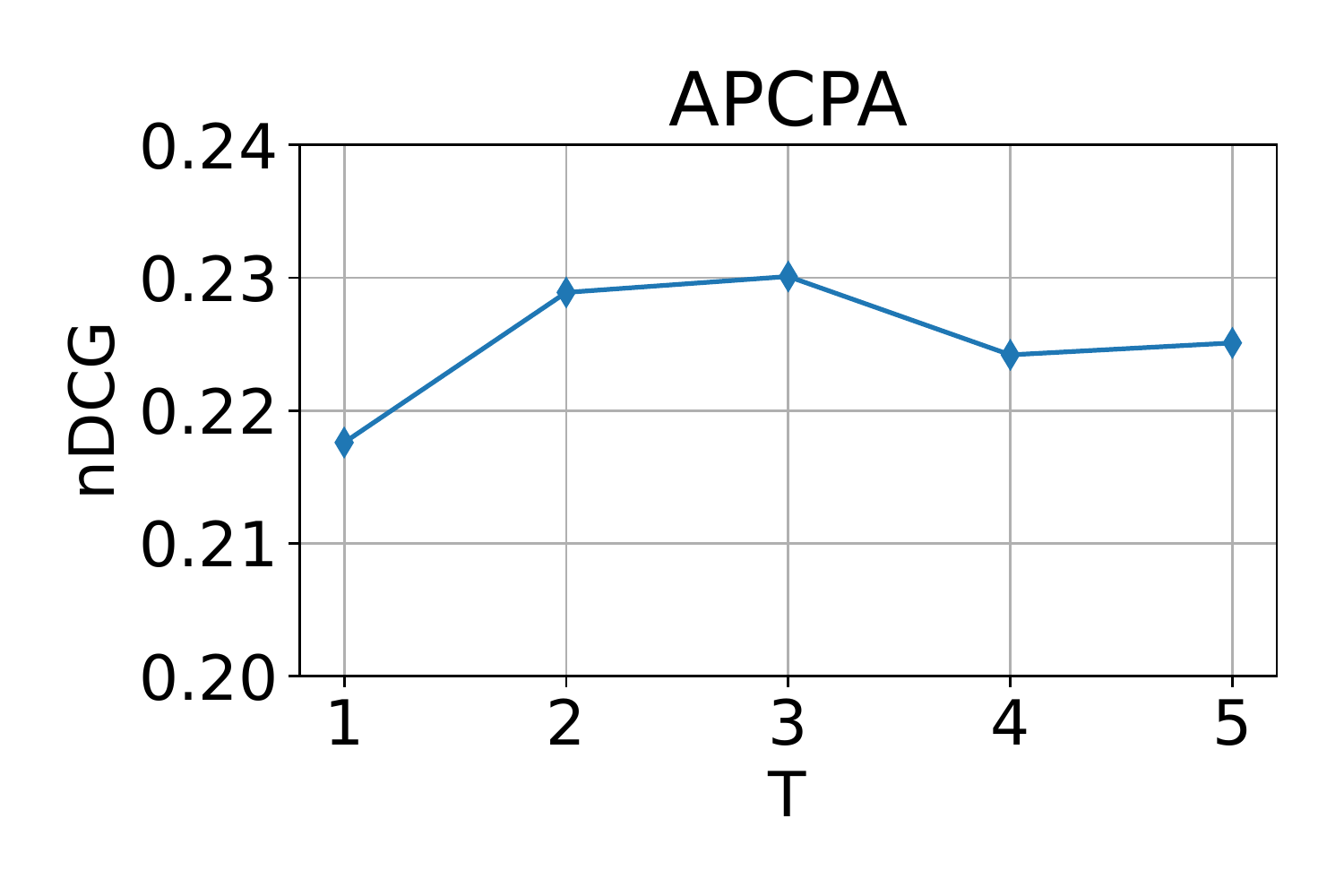}}
    \caption{Performance of NeuPath on varying the $T$ under``APA'' and ``APCPA''.}
    \label{fg:T_ps}
    %\label{fig-online-compare}
\end{figure}

In this section, we evaluate how different choices of $T$, the number of selected path instances, affect the performance of NeuPath on similarity search.  
We test the influence of $T$  with $T = \left \{ 1, 2, 3, 4, 5 \right \}$. The results are shown in Figure \ref{fg:T_ps}. 
We get the best results when $T = 2$ on PathSim under meta-path ``APA'' and $T = 3$ under meta-path ``APCPA''. 
When $T = 1$, it is much worse which proves that using only one path instance is not enough for approximating PathSim. Besides, when $T = 4$ or $T=5$, the performance becomes worse, which shows that the model might select some path instances that do not follow the meta-path in the ground truth PathSim. 

\begin{figure}[t]
\centering
\includegraphics[width=0.40\textwidth]{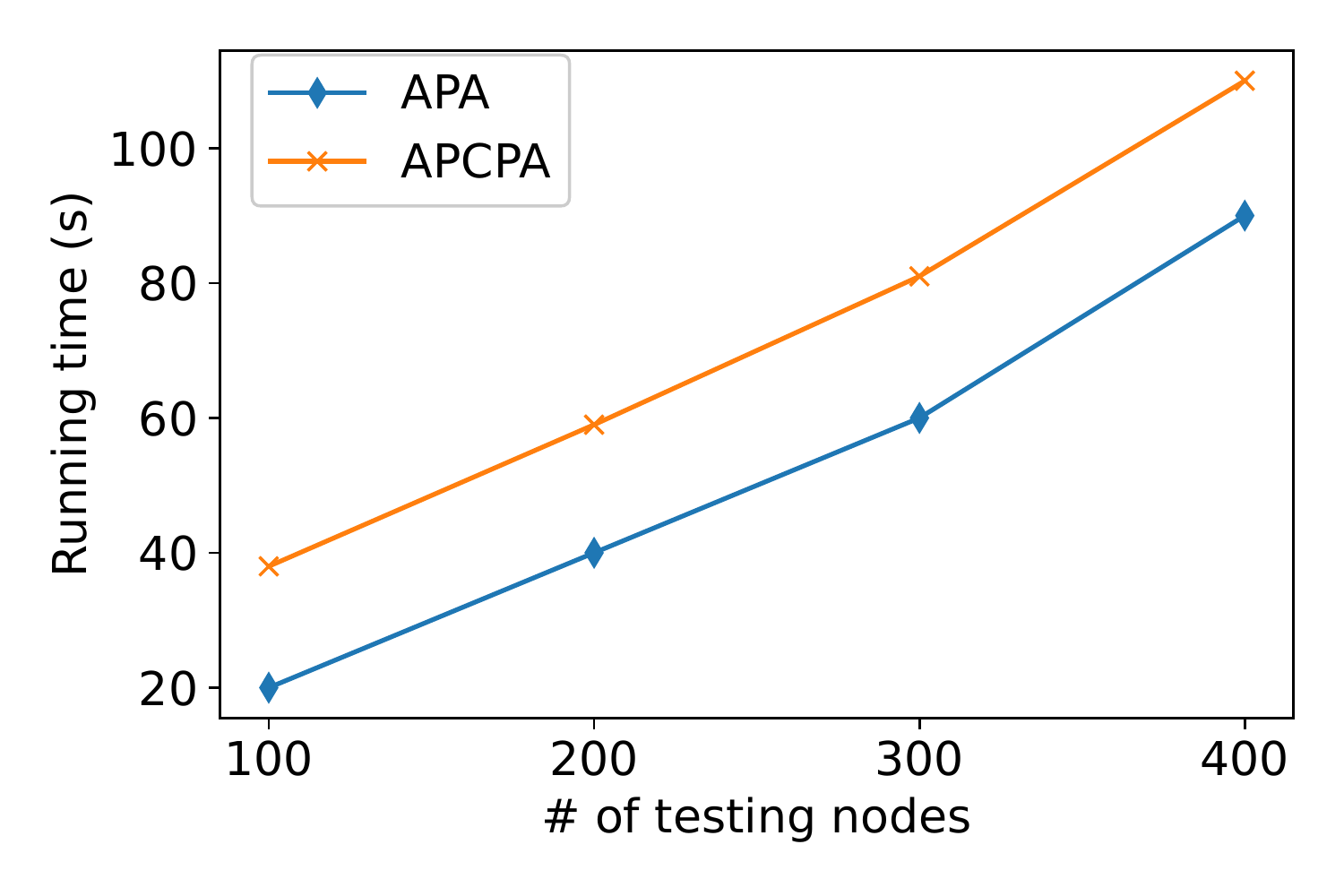}
\vspace{-0.1in}
\caption{Running time with the number of testing nodes growing on testing.} 
\label{fig:running_time}
\end{figure}

\subsection{Analysis of Running Time}
\label{sec:running_time}
% In this section, we show the comparisons of running time between exact PathSim computation and the proposed NeuPath in Table \ref{tb:running time}. We can see that NeuPath runs faster than the PathSim exact method which computes PathSim scores by the definition under meta-paths with different lengths.
As shown in Figure \ref{fig:running_time}, the running time of NeuPath to compute PathSim scores for testing nodes is linear to the number of testing nodes with one GPU device. It is because that NeuPath gains the ability to approximate PathSim for any node in the graph after training the model for one time.

\subsection{Generalization Analysis}
\label{sec:generalization}

\begin{figure}[t]
    \centering
    \subfigure[\% of nodes  under ``APA''.]{\includegraphics[width=0.23\textwidth, height=0.14\textheight]{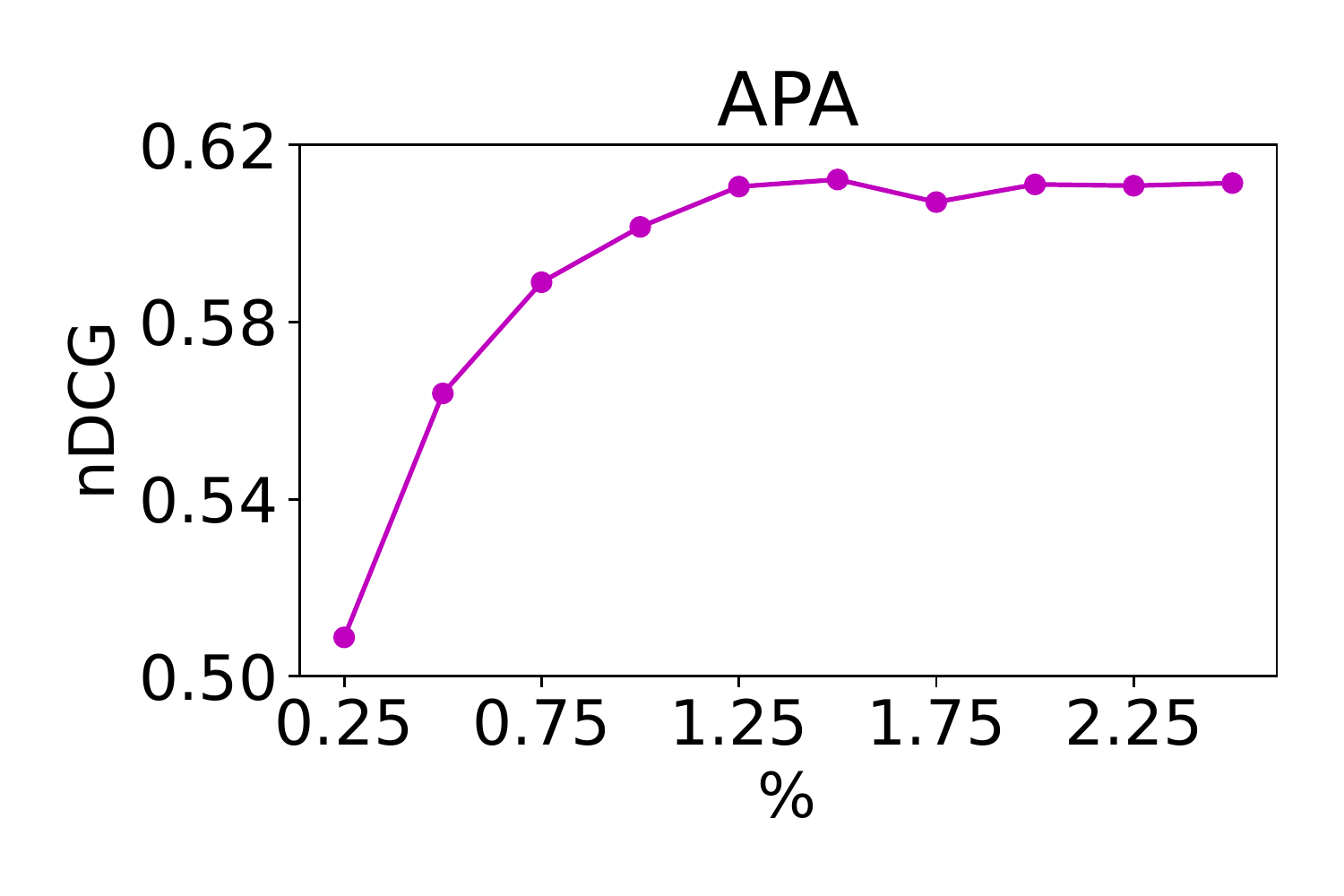}}
    \subfigure[\% of nodes under ``APCPA''.]{\includegraphics[width=0.23\textwidth, height=0.14\textheight]{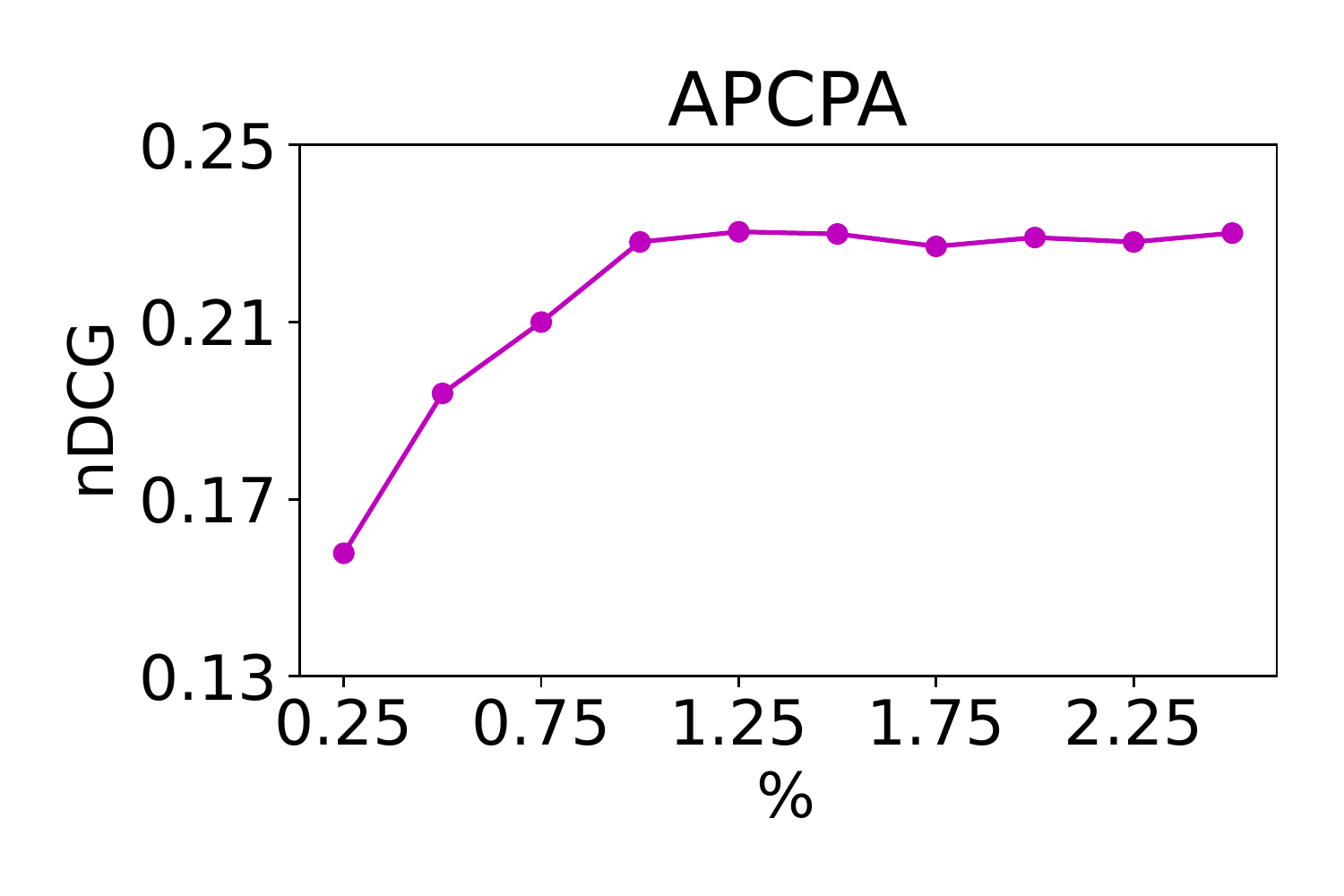}}
    \caption{Performance of NeuPath on varying the percentage of nodes for training under ``APA'' and ``APCPA''.}
    \vspace{-0.1in}
    \label{fg:scalability}
\end{figure}

In this section, we evaluate whether NeuPath could coverage and get the best performance with only a small number of training samples. When applying NeuPath to real-world industry, we manually calculate PathSim scores and get the ranking lists for some query nodes. Then we use these pairwise PathSim scores as the ground truth to train the model, making it be able to learn the strategy of approximating PathSim for any pair of nodes, regarded as inductive ability. Therefore, if the model could learn to approximate with a few training samples, it will have a better inductive ability and be more feasible to datasets with very large scale graphs in real-world applications. We perform NeuPath on varying the percentage of training samples on ACM under meta-path ``APA'' and ``APCPA'', shown in Figure \ref{fg:scalability}. NeuPath could get relatively good results with only 1.25\% training samples. It indicates that NeuPath has good generalization ability and can save a lot of manual labeling for training the model.

\section{Conclusion and Future Work}
\label{sec:conclusion}
%In this paper, we transform the traditional PathSim process for similarity search to a neural version, where an inductive embedding model is learned to search the most helpful path instances for getting the correct PathSim scores. It handles the problem that previous works suffer from computational challenges for the huge graph. Our model, NeuPath is constituted with a neighborhood-aggregation encoder and a multi-layer perceptron decoder,  and it is a novel approach which represents the path instances rather than the nodes embeddings for the similarity calculation. Moreover, NeuPath is able to apply to inductive settings where training and testing are independent. Extensive experiments on real-world datasets ACM and IMDB demonstrate that NeuPath performs better than state-of-the-art learning based baselines in the similarity search task and could approximate the ground truth PathSim scores better than others. Besides, NeuPath could achieve good performance with only a small number of training samples, which proves that NeuPath has a powerful inductive ability and is feasible to large scaled datasets in real-wold industries.

In this paper, we transform the traditional PathSim process for similarity search to a neural version, which handles the problem that previous works suffer from computational challenges for the huge graph. Our model, NeuPath is a novel approach that represents the path instances rather than the nodes embeddings for the PathSim approximation. 
%Moreover, NeuPath can apply to inductive settings where training and testing are independent. 
Extensive experiments on two real-world datasets demonstrate that NeuPath performs better than baselines in the similarity search task and could approximate the ground truth PathSim scores better. In addition, NeuPath could achieve good performance with only a small number of training samples, which proves that NeuPath has a powerful generalization ability and is feasible to large scaled datasets in real-wold industries. In the future work, we will analyze on how to infer the semantics rules between two nodes since NeuPath identifies several helpful path instances.  

% There are several future directions for this work, such as inferring the semantics rules between two nodes since NeuPath identifies several helpful path instances. 
% Also, we can extend the framework to approximate other structural measures, such as SimRank and PageRank.

\section{Acknowledgements}
The authors of this paper were supported the NSFC Fund (U20B2053 ) from the NSFC of China, the RIF (R6020-19 and R6021-20) and the GRF (16211520) from RGC of Hong Kong, the MHKJFS (MHP/001/19) from ITC of Hong Kong, with special thanks to the
HKUST-WeBank Joint Lab at HKUST and their support from the National Key Research and Development Program of China (208AAA0101100).
% Also, we can apply it to some similarity search related applications (e.g. community detection and Web search) and extending the framework to approximate other structural measures, such as geodesic similarity.

\clearpage
\bibliographystyle{plainnat}
\balance
\bibliography{acmart}

\end{document}